\documentclass[conference]{IEEEtran}
\IEEEoverridecommandlockouts
\usepackage{cite}
\usepackage{amsmath,amssymb,amsfonts}
\usepackage{graphicx}
\graphicspath{{figs/}}
\usepackage[caption=false,font=footnotesize]{subfig}
\usepackage{textcomp}
\usepackage{xcolor}
\usepackage{bm}
\usepackage{url}
\usepackage{algorithm}
\usepackage{algorithmicx}
\usepackage{algpseudocode}
\usepackage{cuted}
\usepackage{multirow}

\usepackage{import}

\newcommand{\argmax}{\mathop{\rm arg~max}\limits}

\def\BibTeX{{\rm B\kern-.05em{\sc i\kern-.025em b}\kern-.08em
    T\kern-.1667em\lower.7ex\hbox{E}\kern-.125emX}}
\begin{document}

\title{Scalable Base Station Configuration via Bayesian Optimization with Block Coordinate Descent
\thanks{This work was supported by JST PRESTO under Grant Number JPMJPR23P3, JST ASPIRE under Grant JPMJAP2346, and JSPS KAKENHI Grant Number 	25K00138.}}

\author{\IEEEauthorblockN{Kakeru Takamori and Koya Sato}
\IEEEauthorblockA{Artificial Intelligence eXploration Research Center,\\
The University of Electro-Communications, 1-5-1, Chofugaoka, Chofu-shi, Tokyo, Japan \\
E-mail: k.takamori@uec.ac.jp, k\_sato@ieee.org}
}

\maketitle

\begin{abstract}
    This paper proposes a scalable Bayesian optimization (BO) framework for dense base-station (BS) configuration design. BO can find an optimal BS configuration by iterating parameter search, channel simulation, and probabilistic modeling of the objective function. However, its performance is severely affected by the curse of dimensionality, thereby reducing its scalability. To overcome this limitation, the proposed method sequentially optimizes per-BS parameters based on block coordinate descent while fixing the remaining BS configurations, thereby reducing the effective dimensionality of each optimization step. Numerical results demonstrate that the proposed approach significantly outperforms naive optimization in dense deployment scenarios.
\end{abstract}


\section{Introduction}
\label{sect:intro}
The rapid growth of wireless devices driven by the Internet of Things (IoT) and sixth-generation (6G) services has significantly increased wireless traffic demand, making careful \emph{a priori} design of multi--base-station (BS) systems indispensable.
In particular, BS locations, transmission powers, and antenna configurations must be optimized prior to deployment to maximize area-wide performance metrics such as the average throughput.
\par
A key challenge in BS design lies in site-specific propagation effects that cannot be accurately captured by simple path-loss models and must instead be evaluated through high-fidelity channel simulations, resulting in computationally expensive trial-and-error design processes.
This motivates the development of algorithms that can efficiently explore BS configurations with a limited number of evaluations.
\par
Bayesian optimization (BO) has emerged as a promising black-box optimization framework for such simulation-based design problems, especially when optimizing spatially averaged performance metrics, e.g., area-wide throughput~\cite{k_sato}.
However, the performance of BO degrades rapidly as the dimensionality of the input space increases.
In multi-BS design, the dimensionality scales with the number of BSs, since each BS introduces multiple design variables, which makes naive BO impractical for dense deployments.
\par
To address this issue, this work proposes a scalable BO-based framework for dense BS configuration design by integrating block coordinate descent (BCD)~\cite{Wang2023-nx}.
By decomposing the high-dimensional optimization problem into a sequence of per-BS subproblems and optimizing them sequentially while fixing the others, the proposed method significantly improves the scalability of BO for dense BS deployments.
\section{System Model}
\label{sect:systemmodel}
We consider a three-dimensional space $\mathcal{A}$ in which $N_{\mathrm{Tx}}$ base stations (BSs) operate on a channel.
The set of BS parameters is denoted by $\Phi=\{\phi_1,\phi_2,\cdots,\phi_{N_{\mathrm{Tx}}}\}$, where the parameters in the $i$-th BS is $\phi_i=\{\mathbf{x}_i,P_i,\bm{\theta}_i\}$.
$\mathbf{x}_i$ is the placement location, $P_i$ is the transmission power, and $\bm{\theta}_i$ specifies its antenna orientation.
The parameter $\bm{\theta}_i$ comprises three angles (yaw, pitch, roll), which represent the azimuth angle, elevation angle, and left--right tilt, respectively.
\par
Let the channel bandwidth be $B\,[\mathrm{Hz}]$.
For a receiver at location $\mathbf{x}$, let $\mathrm{SINR}_i(\mathbf{x})$ denote the SINR associated with the $i$-th BS.
The objective is to find the parameter set $\Phi^{\mathrm{opt}}$ that maximizes the area-averaged throughput, where the throughput from the $i$-th BS is given by $C_i(\mathbf{x})=B\log_2(1+\mathrm{SINR}_i(\mathbf{x}))$.
This can be formulated as
\begin{equation}
    \Phi^{\mathrm{opt}}=\argmax_{\Phi}  \left[f(\Phi)\triangleq\frac{1}{|\mathcal{A}|}\sum_i \int_\mathcal{A} C_i(\mathbf{x})\,d\mathbf{x}\;[\mathrm{bps}/\mathrm{m}^2]\right], \nonumber
\end{equation}
where $|\mathcal{A}|$ denotes the total area of $\mathcal{A}$.

\section{Scalable BS Configuration via BCD-Aided BO}
\label{sect:proposed}

\begin{figure}[t]
  \centering
  \includegraphics[width=\linewidth]{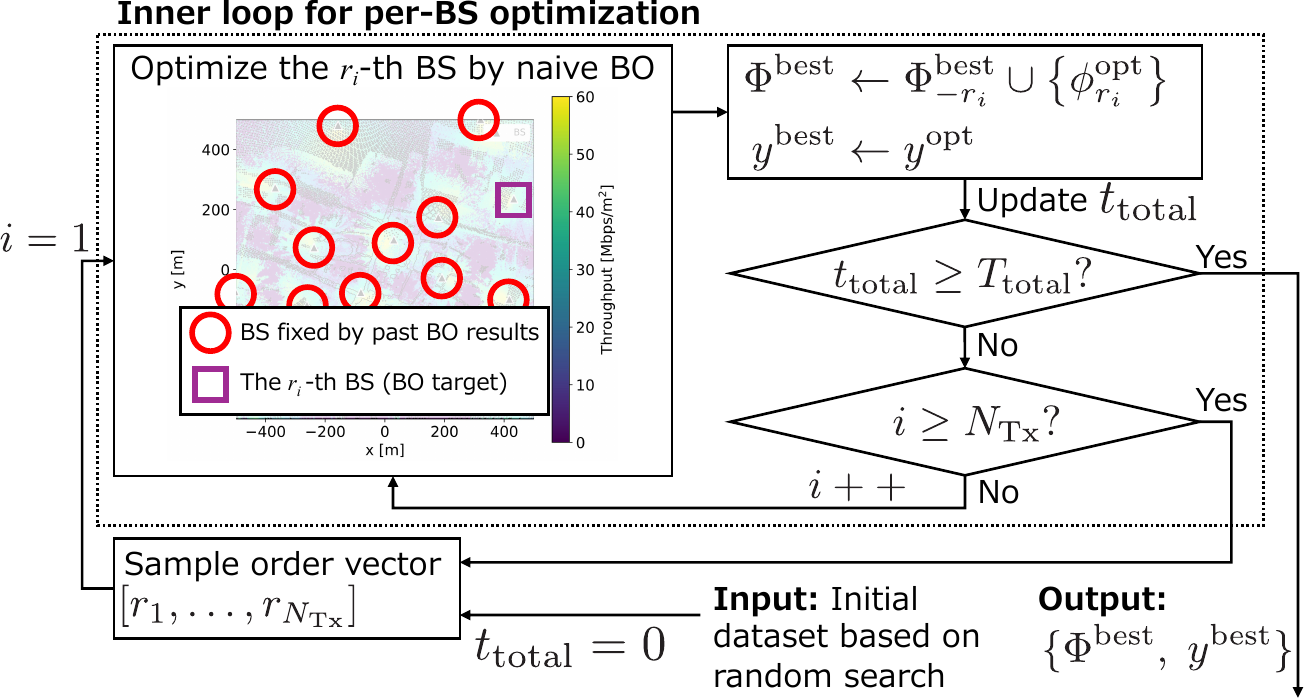}
  \caption{Overview of the proposed BCD-aided BO.}
  \label{fig:opt_flow}
  \vspace{-2mm}
\end{figure}

Fig.\,\ref{fig:opt_flow} summarizes the proposed BCD-aided BO.
BO can explores BS configurations by iterating parameter selection, ray-tracing–based performance evaluation, and probabilistic modeling of evaluated configurations. However, a naive BO approach that jointly optimizes all parameters in $\Phi$ suffers from the curse of dimensionality, resulting in poor convergence and a large number of costly ray-tracing evaluations, since the parameter space scales as $6\times N_{\mathrm{Tx}}$.
To overcome this limitation, we integrate BCD into BO, reducing the effective dimensionality by optimizing one variable block at a time while fixing the others.
In multi-BS design, the parameter set of each BS forms a block; exploiting this structure, we sequentially optimize each BS by BO.
All updates over the BSs are regarded as one cycle, and such cycles are repeated to progressively refine the overall configuration.
\par
The proposed method consists of outer and inner loops, as shown in Fig.\,\ref{fig:opt_flow}.
Each outer loop randomly samples a permutation of BS indices $[r_1,\ldots,r_{N_{\mathrm{Tx}}}]$ to determine the update order of BSs; subsequently, in the inner loop, BO is applied to each BS individually, and each BS is updated sequentially.
Let $\Phi^{\mathrm{best}}$ denote the best BS configuration obtained so far, and let $y^{\mathrm{best}}$ be the corresponding objective value.
For the update of the $r_i$-th BS, we optimize only $\boldsymbol{\phi}_{r_i}$ while fixing the remaining parameters $\Phi^{\mathrm{best}}_{-r_i}\leftarrow \Phi^{\mathrm{best}}\backslash \{\phi_{r_i}\}$; the subproblem for per-BS update can be written as
\begin{equation}
    \max_{\phi_{r_i}} \left[g_{r_i}(\phi_{r_i}) \triangleq f\left(\Phi \mid \Phi^{\mathrm{best}}_{-r_i}\right)\right].\nonumber
\end{equation}
We model $g_{r_i}(\phi_{r_i})$ by Gaussian process with a radial basis function kernel and select the next experimental parameter based on maximizing the expected improvement.
This subproblem is solved based on a naive BO: we initialize the observation set by including $\{\phi_{r_i}^{\mathrm{best}},y^{\mathrm{best}}\}$ and $N_{\mathrm{init}}$ random samples, and then iteratively select and evaluate candidates until $t_{\mathrm{sub}}$ reaches $T_{\mathrm{sub}}$.
After that, we update the incumbent as
$\Phi^{\mathrm{best}}\leftarrow \Phi^{\mathrm{best}}_{-r_i} \cup \left\{\phi_{r_i}^{\mathrm{opt}}\right\}$, $y^{\mathrm{best}}\leftarrow y^{\mathrm{opt}},$
where $\{\phi_{r_i}^{\mathrm{opt}},y^{\mathrm{opt}}\}$ is the best observed pair in the BO run for the $r_i$-th BS.
Because $g_{r_i}$ depends on the fixed context $\Phi^{\mathrm{best}}_{-r_i}$, the observation set is re-initialized for each BS update.
We repeat the above process until $t_{\mathrm{total}}$ reaches $T_{\mathrm{total}}$; finally, we set $\Phi^\mathrm{opt}\leftarrow \Phi^\mathrm{best}$.
\section{Numerical Results}
\label{sect:performance}
\begin{figure}[t]
  \centering
  \setcounter{subfigure}{0}

  \subfloat[]{%
    \includegraphics[width=0.48\linewidth]{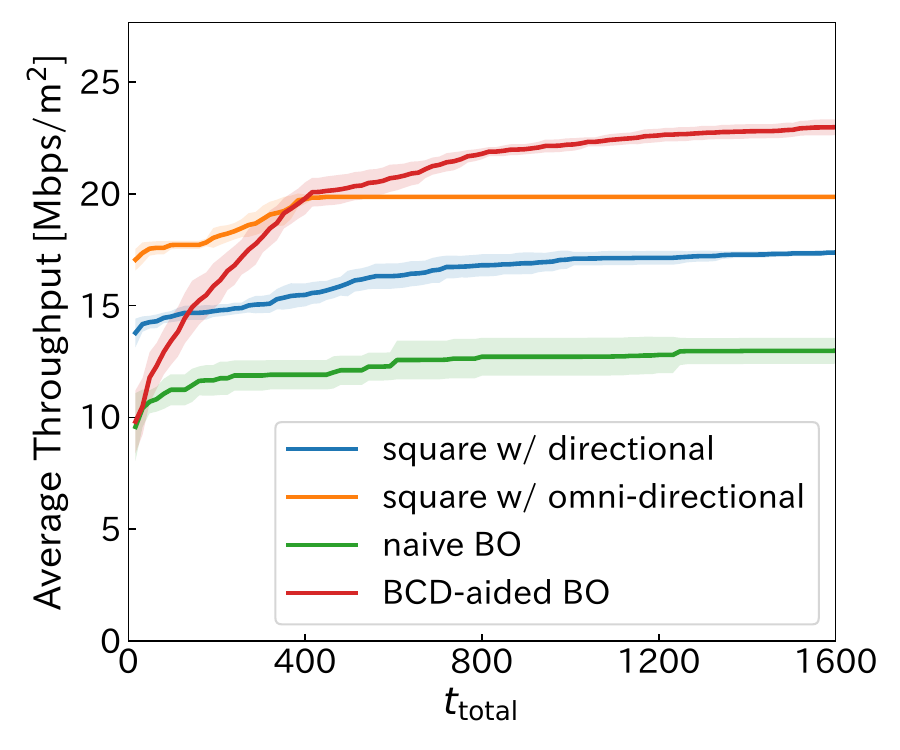}%
    \label{fig:ntx16}
  }\hfill
  \subfloat[]{%
    \includegraphics[width=0.48\linewidth]{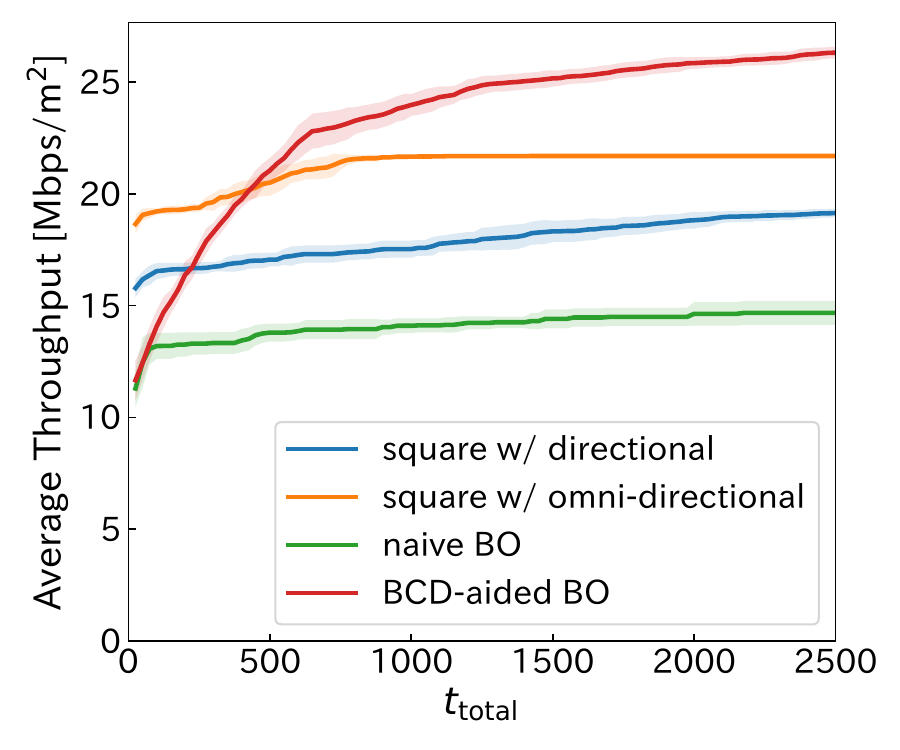}%
    \label{fig:ntx25}
  }

  \caption{Throughput versus $t_{\mathrm{total}}$: (a)\,$N_\mathrm{Tx}=16$, (b)\,$N_\mathrm{Tx}=25$.}
  \label{fig:regret}
  \vspace{-2mm}
\end{figure}

We evaluate the proposed method using Sionna RT\footnote{\url{https://nvlabs.github.io/sionna/rt/index.html}}.
The evaluation area is \texttt{munich}, a 3D urban scene provided by Sionna RT, and we focus on the central $1\,\mathrm{km}\times 1\,\mathrm{km}$ region.
The BS antenna height is $20\,\mathrm{m}$ and the receiver antenna height is $1.5\,\mathrm{m}$.
We compare the four approaches: (a)\,square placement with directional antenna (with BO-based optimization for power and orientation), (b)\,square placement with omni-directional antenna (with power optimization), (c)\,naive BO, and (d)\,the proposed method (BCD-aided BO).
The evaluation budget is $T_{\mathrm{total}}=100\times N_{\mathrm{Tx}}$; for the proposed method, $T_{\mathrm{sub}}=25$ and $N_{\mathrm{init}}=10$.

\begin{figure}[t]
\centering
\setcounter{subfigure}{0}
\newlength{\OuterColW}
\setlength{\OuterColW}{\columnwidth}

\newcommand{\ShiftLeft}{-0.01\OuterColW}
\newcommand{\MapBlockW}{0.83\OuterColW}
\newcommand{\CbBlockW}{0.04\OuterColW}
\newcommand{\GapMC}{0.02\OuterColW}
\newcommand{\EachMapW}{0.495\linewidth}
\newcommand{\CbH}{0.83\OuterColW}
\newcommand{\RowGap}{-1.2ex}

\makebox[\OuterColW][l]{\hspace*{\ShiftLeft}%
  \begin{minipage}[t]{\MapBlockW}
    \centering
    \vspace{0pt}

    \subfloat[]{%
      \includegraphics[width=\EachMapW]{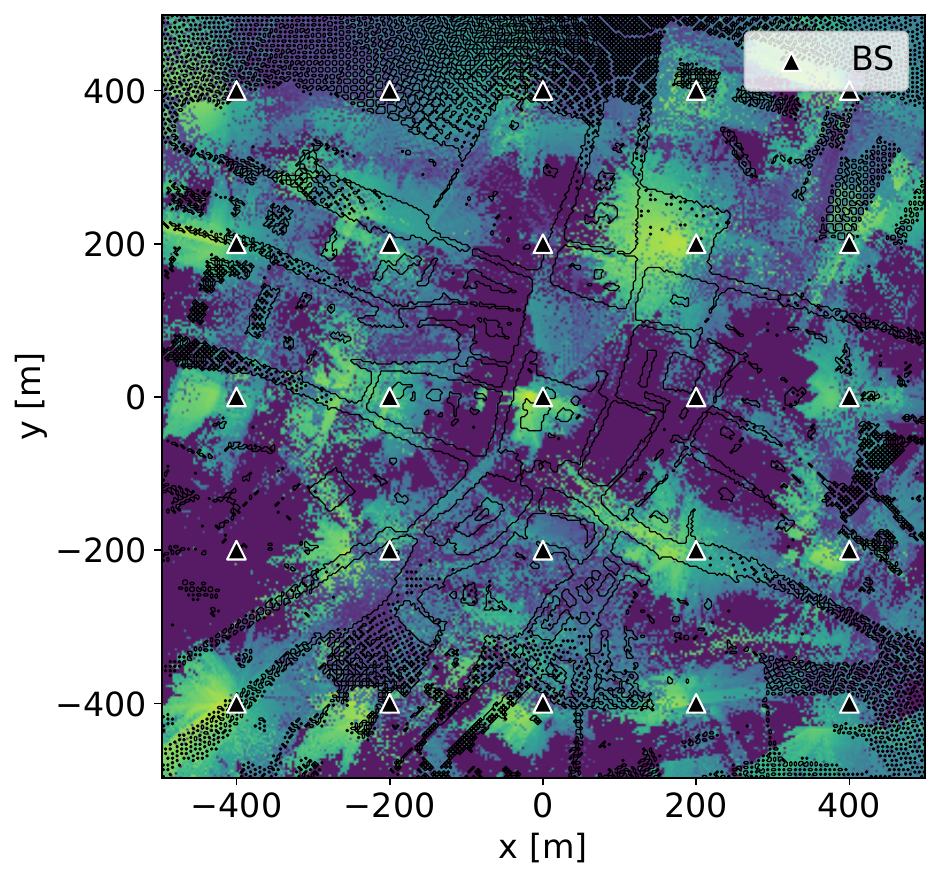}%
    }\hfill
    \subfloat[]{%
      \includegraphics[width=\EachMapW]{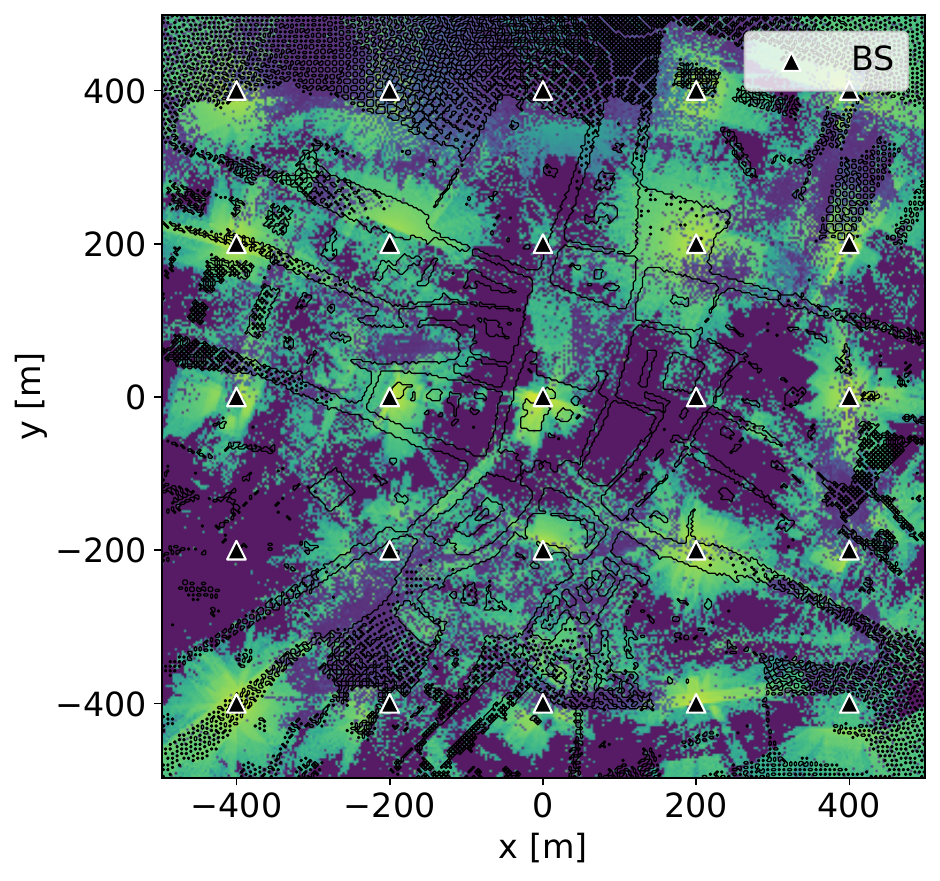}%
    }\\[\RowGap]

    \subfloat[]{%
      \includegraphics[width=\EachMapW]{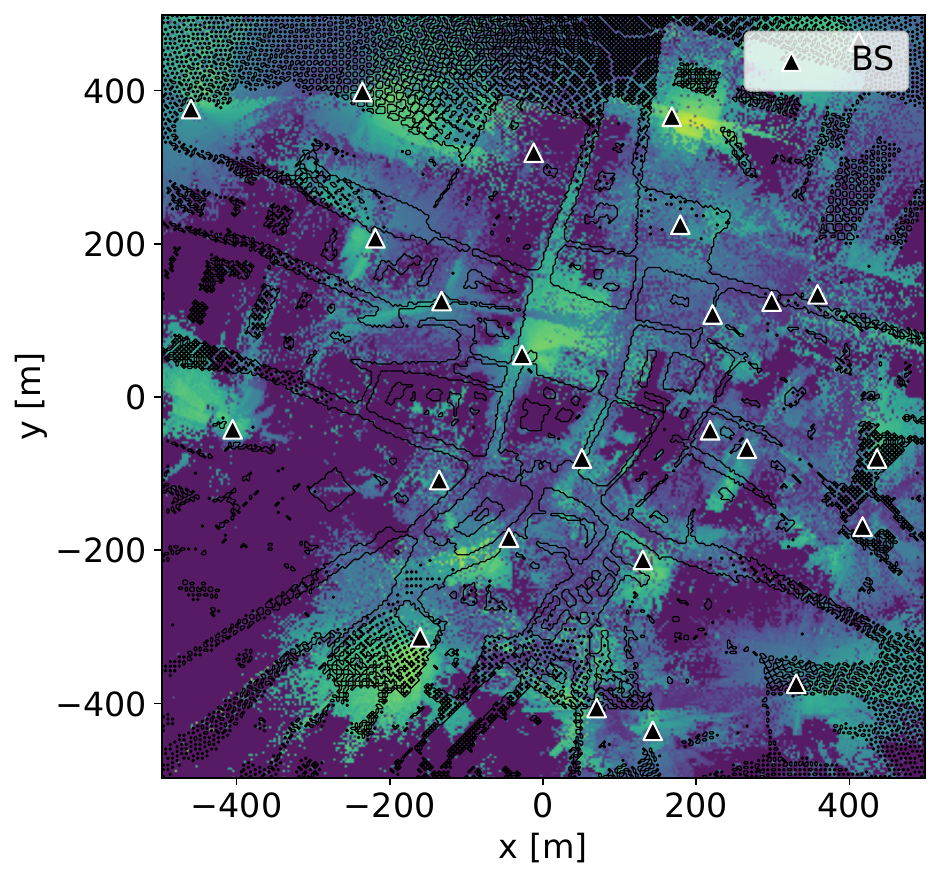}%
    }\hfill
    \subfloat[]{%
      \includegraphics[width=\EachMapW]{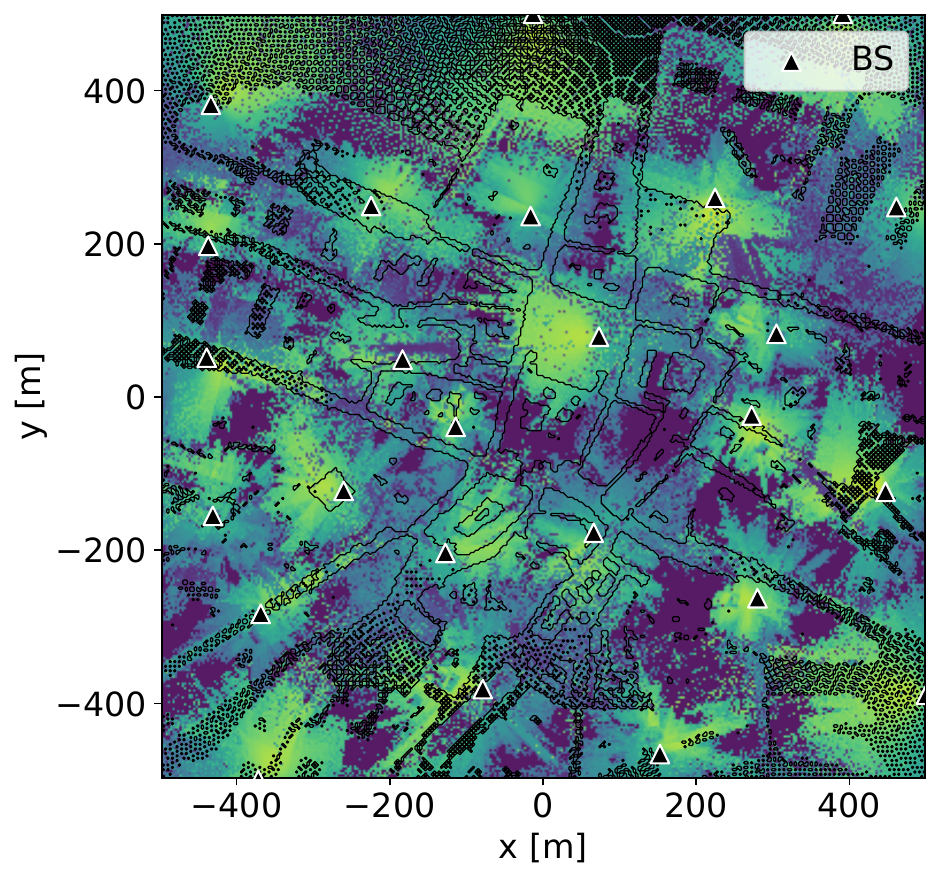}%
    }

  \end{minipage}%
  \hspace{\GapMC}%
  \begin{minipage}[t]{\CbBlockW}
    \centering
    \vspace{7.05pt}
    \includegraphics[height=\CbH]{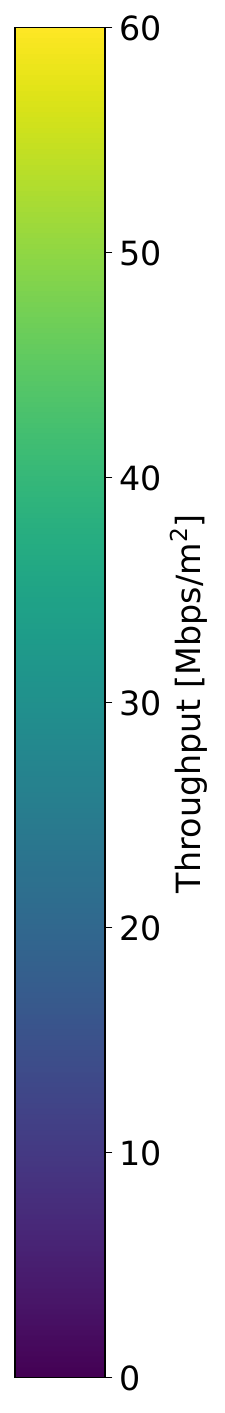}
  \end{minipage}%
}

\caption{Optimization examples at $N_{\mathrm{Tx}}=25$ and $t_{\mathrm{total}}=2500$: (a)\,square placement with directional antennas, (b)\,square placement with omni-directional antennas, (c)\,naive BO, and (d)\,proposed BCD-aided BO.}
\label{fig:heatmap_best_25}
\vspace{-2mm}
\end{figure}

Fig.~\ref{fig:regret} shows the best-so-far area-averaged throughput as a function of $t_{\mathrm{total}}$.
Our method demonstrates a substantial performance gain over baseline approaches; for example, the proposed method achieves 15.8\% and 21.5\% higher area-averaged throughput than the omni-directional square-placement for $N_{\mathrm{Tx}}=16$ ($t_\mathrm{total}=1600$) and $25$ ($t_\mathrm{total}=2500$), respectively.
Further, Fig.~\ref{fig:heatmap_best_25} visualizes the optimization examples for $N_{\mathrm{Tx}}=25$.
These results indicate that the proposed method can achieve high throughput performance across a wide area.

\section{Conclusion}
We studied BS-parameter optimization for dense deployments and proposed a per-BS sequential search method that integrates BCD with BO.
Numerical results showed significant gains in area-averaged throughput compared with related methods.
Future work includes validation on additional urban scenes and optimization of the BS update order.

\bibliographystyle{IEEEbib}
\bibliography{reference}

\end{document}